\documentclass[conference]{IEEEtran}

\usepackage{psfig}
\usepackage{array}
\usepackage{epsfig}
\usepackage{amsmath}
\usepackage{amssymb}
\usepackage{graphicx}
\usepackage{amsthm}

\newtheorem{theorem}	{Theorem}

\begin{document}

\title{Randomly Roving Agents in Wireless Sensor Networks}

\author{\IEEEauthorblockN{Hakob Aslanyan}
\IEEEauthorblockA{Computer Science Department\\University of Geneva\\
1227 Geneva, Switzerland\\
hakob.aslanyan@unige.ch}
\and
\IEEEauthorblockN{Jose Rolim}
\IEEEauthorblockA{Computer Science Department\\University of Geneva\\
1227 Geneva, Switzerland\\
jose.rolim@unige.ch}}

\maketitle

\begin{abstract}
Quantitative characterization of randomly roving agents in wireless sensor networks (WSN) is studied. Below the formula simplifications, regarding the known results and publications, it is shown that the basic agent model is probabilistically equivalent to a similar simpler model and then a formula for frequencies is achieved in terms of combinatorial second kind Stirling numbers. Stirling numbers are well studied and different estimates are known for them letting to justify the roving agents quantitative characteristics.
\end{abstract}

\IEEEpeerreviewmaketitle

\section{Introduction}
This work, inspired by \cite{krugelmobileagent,irarandomagents,brandesnetanalysis}, considers roving agents' numerical characterization, challenging ad hoc pervasive and trustworthy networks. Agents are autonomous, moving, and intelligent software structures capable to play a sensitive role in advanced monitoring, computation and protection systems. Intrusion detection systems (IDS) \cite{krugelmobileagent} are addressed particularly. They appear as complementary mean to the ordinary cryptographic protection tools of computers and networks. Such IDS use software agent based monitoring and data collection, watching the inside processes of a computer, registering LOG files of application software systems, sniffing and recording communication protocols. Watching the whole network behavior they are better suited to warn approaching attacks and malfunctioning. Data mining agents (DMA) and Data fusion agents (DFA) are examples of information integration tools in networks \cite{irarandomagents}. In large networks, moreover when its structure is not predefined such as wireless sensor networks \cite{brandesnetanalysis} it is natural to consider independent, randomly roving agents, requiring that they are able to collect enough information in total, mining the necessary knowledge about the intrusion. This framework is studied in \cite{irarandomagents}, which prove formulas for the number of DMA sufficient to monitor the given size areas of networks. The formula received is complex and impractical because of their use of nested sums by different parameters. Our work tends to prove simple estimates for the same numerical characteristics of WSN.

\section{Roving Agents Model}
DMA roams around randomly in a network and acquires environmental information. It is lightweight using simplest mining algorithms. DFA is for integration of DMA set actions. DFA may act as an intrusion detection tool and then its power depends on information collected by DMA in network.

Let we are given a network $N$ of $n$ nodes $v_{1},\ldots,v_{n}$. Some fixed amount of information $\theta_{i}$ is allocated at node $v_{i}$. There are $k$ DMA  $a_{1},\ldots,a_{k}$. Each agent \textit{\textbf{visits exactly $m$ different nodes}} and obtains the unique information content in each such node. DMA pass all collected information to DFA. Denote by $P_{k}(n,m,t)$ the probability that DFA contains exactly $t$ information blocks of network nodes when $k$ agents randomly visit $m$ of $n$ nodes each. The formula for $P_{k}(n,m,t)$ proven in \cite{irarandomagents} looks as:
\begin{align}
 P_{k}  &(n,m,t)= \nonumber \\ &\binom{n}{m}^{-(k-1)}\sum_{m_{2},m_{3},\ldots,m_{k-1}=0}^{m}\binom{m}{m_{2}} \binom{n-m}{m-m_{2}} \cdot \nonumber \\
 \cdot  &\binom{2m-m_{2}}{m_{3}}\binom{n-2m+m_{2}}{m-m_{3}}\ldots \nonumber \\
 \ldots &\binom{(k-2)m-m_{2}-\ldots-m_{k-2}}{m_{k-1}} \cdot \nonumber \\
 \cdot  &\binom{n-(k-2)m+m_{2}+\ldots+m_{k-2}}{m-m_{k-1}}\cdot \nonumber \\ 
 \cdot  &\binom{(k-1)m-m_{2}-\ldots-m_{k-1}}{km-t-m_{2}-\ldots-m_{k-1}}\cdot \nonumber \\
 \cdot  &\binom{n-(k-1)m+m_{2}+\ldots+m_{k-1}}{t-(k-1)m+m_{2}+\ldots+m_{k-1}},k\geq4. \label{pbigformula}
\end{align}

Formulas for smaller $k$ given in \cite{irarandomagents} look similar to $(\ref{pbigformula})$. Of course these formulas are unobservable and simplifications or approximations are of interest. By this same reason \cite{irarandomagents} proves formulas, considering computer simulation, to understand the typical numbers of agents necessary to retrieve the required information in network. Modifications of ``\textit{exactly} $t$'' condition in agent distribution scheme are also important to be considered.

\section{Coverage Characterization of Roving Agents}
Let we are given the set $N=\{v_{1},\ldots,v_{n}\}$ of nodes and $S_{1},\ldots,S_{k}$ are $k$ arbitrary subsets of $N$, each of size $m\leq n$, visited correspondingly by the $k$ agents. We consider a probability distribution scheme over the $N$, and suppose that  $m$-subsets $S_{j}$ are equiprobable and independent in this scheme. Having in total $C_{n}^{m}$ $m$-subsets the probability of one of them is equal to $1/C_{n}^{m}$. We are interested in knowing the probabilistic characteristics of the union $\cup_{i=1}^{k}S_{i}$ and its size. In particular, what is the probability that union of those subsets contains exactly $t$ elements?
\begin{align}
P_{k}(n,m,t)=Pr\left(\left|\bigcup_{i=1}^{k}S_{i}\right|=t\right). \label{pt}
\end{align}

To collection of subsets $S_{1},\ldots,S_{k}$ of $N$ nodes corresponds a matrix $A^{k \times n}=\{a_{ij}\}$ where
\begin{align}
 a_{ij} = 
  \begin{cases} 
   1 & \text{if } v_{j}\in S_{i} \\
   0 & \text{otherwise}
  \end{cases}.
\end{align}

As each $S_{i}$ contains exactly $m$ elements then each row of $A^{k \times n}$ will contain $m$ $1$s and $n-m$ $0$s. If $\left|\cup_{i=1}^{k}S_{i}\right|=t$, then there are $t$ columns of $A$ which contain at least one $1$ and $n-t$ columns which don't contain $1$. The number of $k \times n$ matrixes with $m$ ones on each row and with exactly $n-t$ columns with no $1$ is $C_{n}^{t} \cdot Q(k,m,t)$ where $Q(k,m,t)$ is the number of $k \times t$ matrixes with $m$ ones on each row and at least one $1$ on each column. 

\begin{figure}[tb]
\begin{center}
\begin{minipage}[h]{\linewidth}  
\includegraphics[width=.8\textwidth]{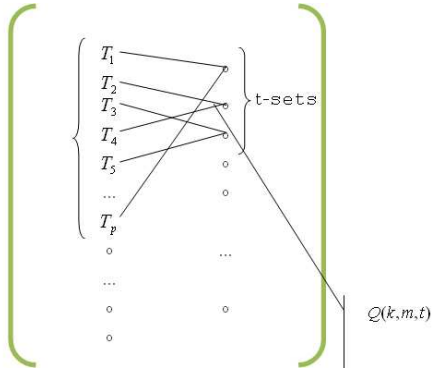}
\end{minipage}
\end{center}
\caption{Agent sets distribution in terms of trials and node sets. Left column contains outcomes of $k$ by $m$ trials (each $T_{i}$ is a ordered collection of $k$ $m$-subsets). Right column contains all the subsets of node set $N$.}
\label{fig:distributionk_m}
\end{figure}

Alternatively, let us consider the following schematic presentation of roving agents' distribution. 
Left column vertices in the scheme presented in Fig. \ref{fig:distributionk_m} contain all the arrangements $T_{1},T_{2},\ldots$ of $k$ agents roving by $C_{n}^{m}$ $m$-node-subsets (ordered collections of $k$ $m$-node-subsets).
From combinatorial perspective agents and nodes are distinguishable but $m$-node-subsets are considered as usual sets - different elements and no ordering. Total number of arrangements is equal to $\left(C_{n}^{m}\right)^{k}$. Part of these arrangements cover exactly $t$ nodes and let that these are vertices $T_{1},T_{2},\ldots,T_{p}$. In this notation $p$ is the unknown number that we want to compute. Right side column vertices correspond to all subsets of node set $N$ and part of these sets are of size $t$. In principle, node subset sizes may vary from $0$ to $n$ but in our experiment it may take values from $m$ to $\min(km,n)$.

We draw an edge between an arrangement and a node subset which is covered by that arrangement. Each arrangement is incident to exactly one edge (and subset). Each $t$-subset appears in different arrangements and this number is common for all  $t$-subsets and is given by $Q(k,m,t)$.

$Q(k,m,t)$ can be calculated by inclusion-exclusion principle. We use the matrix model for arrangements. First, over a $k \times t$ matrix we take the whole set of unconstrained arrangements as all matrices with $m$ $1$s on rows, then we remove from this all the arrangements where at least one column is initially filled with $0$ (such matrices do not obey the conditions we require), then add arrangements with at least $2$ empty columns, etc. The formula representation of related quantities is:
\begin{align}
Q(k,m,t) & = \nonumber \\ 
& \left(C_{t}^{m}\right)^{k}-C_{t}^{1} \cdot \left(C_{t-1}^{m}\right)^{k}+C_{t}^{2} \cdot \left(C_{t-2}^{m}\right)^{k}-\ldots \nonumber \\ &+\left(-1\right)^{t-m}C_{t}^{t-m} \cdot \left(C_{m}^{m}\right)^{k}= \nonumber \\ 
& \sum_{i=0}^{t-m}\left(-1\right)^{i}C_{t}^{i} \cdot \left(C_{t-i}^{m}\right)^{k}. \label{inclexcl}
\end{align}

We have proven
\begin{theorem}
\begin{align}
	P_{k}(n,m,t)=\frac{C_{n}^{t} \cdot \sum_{i=0}^{t-m}\left(-1\right)^{i}C_{t}^{i} \cdot \left(C_{t-i}^{m}\right)^{k}}{\left(C_{m}^{n}\right)^{k}}. \label{pformula}
\end{align}
\end{theorem}

First of all here we receive a real simplification of $(\ref{pbigformula})$. The formula received is still complex, but it might be approximated and the applied Markov inequality may give asymptotic estimates of  $t$-subset probabilities \cite{medvedev}.

Another important characteristic, the mean value of subset size $t$, might be computed as:
\begin{align}
	&\sum_{t=m}^{\min(km,n)}t \cdot P_{k}(n,m,t)= \nonumber \\
	&=\sum_{t=m}^{\min(km,n)}\frac{t \cdot C_{n}^{t} \cdot \sum_{i=0}^{t-m}\left(-1\right)^{i}C_{t}^{i} \cdot \left(C_{t-i}^{m}\right)^{k}}{\left(C_{m}^{n}\right)^{k}}. \label{meanformula}
\end{align}

\section{On Node Repetition Limitations in An Agent Roving Scheme}

Let us consider the scene of random distribution of $m$ agents over the $n$ WSN nodes (here we do not consider $k$ agents but $m$ agents, and each individual agent visits exactly one node). Agents are dropped over the node set one by one, independently, and with equal probabilities for nodes. Allocating all $m$ agents we receive a collection of nodes visited by agents, probably with multiple agents that visited the same node.

Total number of different allocations is $n^{m}$. Among these are $1$ node allocations (all the agents visit the same node), their number is $n$, $2$ node allocations, they are $C_{n}^{2}\left(2^{m}-2\right)$ and the largest are $m$ node allocations ($m$-sets), when agents are distributed in all different nodes, and they are $n(n-1)\ldots (n-m+1)$. We are interested in the frequencies of allocation sizes when at least $2$ agents are allocated at the same node (sizes from $1$ to $m-1$), or complementary, the share of allocations with all different nodes.

\begin{figure}[tb]
\begin{center}
\begin{minipage}[h]{\linewidth}
\includegraphics[width=.8\textwidth]{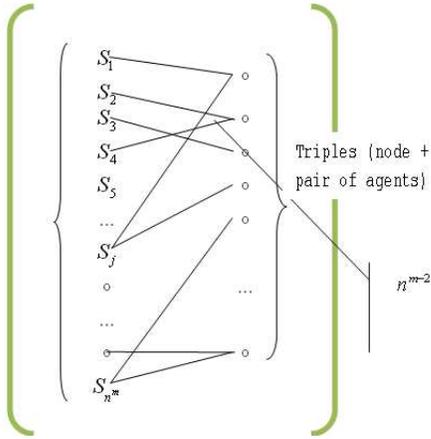}
\end{minipage}
\end{center}
\caption{Agents distribution on WSN node sets. Left column contains outcomes of $m$ trials (each $S_{i}$ is a ordered collection of $m$ nodes), right column contains triples, node and two different agents}
\label{fig:distributionkm}
\end{figure}

One of the classical approaches of determining typical cases in distributions is when Markov or Chebyshev inequality is applied. In this way we consider the scheme presented in Fig. \ref{fig:distributionkm} similar to one presneted in Fig. \ref{fig:distributionk_m} to compute the mean of the number of allocated nodes in random distribution of $m$ agents over the $n$ WSN nodes.

Thus, the number of right side vertices in the scheme, where each vertex is a triple, node and a pair of agents, is $nC_{m}^{2}$. Edges are connecting an allocation (from left column) to a node with the given pair of agents it contains (right column). We compute the mean number $M(v_{n,m})$ of edges incident to each allocation as

\begin{align}
M(v_{n,m})=\frac{nC_{m}^{2} \cdot n^{m-2}}{n^{m}}=\frac{C_{m}^{2}}{n}
\end{align}

Apply Markov inequality $Pr\left\{v_{n,m} \geq \epsilon \right\} \leq M(v_{n,m}) / \epsilon$. Take $\epsilon=1$, then $C_{m}^{2}/n$ is the upper estimate of probability of repeating agents at nodes. If $C_{m}^{2}/n \rightarrow 0$ with $n,m \rightarrow \infty$, then we receive that almost all allocations consist of all different agents at nodes.
\section{Comparison of Agent Allocation Schemes}
In this point we will define and consider two basic probability distributions tightly related to each other.
\begin{itemize}
	\item First distribution $U_{n,k,\left\{m\right\}}$ is composed by $k$ independent consecutive allocations of $m$-node subsets over the WSN area of $n$ nodes. $\left(C_{n}^{m}\right)^{k}$ Outcomes of trials are ordered collections of $m$-subsets of WSN nodes. These collections may cover all node subsets of sizes from $m$ to $\min(km,n)$. 
	
	\item Second distribution scheme $U_{n,k,m}$, which we want to consider and compare with the basic distribution $U_{n,k,\left\{m\right\}}$ considered above, consists of $k$ consecutive and independent stages; each stage allocates $m$ elements consecutively and independently over the WSN area of $n$ nodes. Outcomes of these trials are all $n^{km}$ ordered collections of nodes. These collections may cover all node subsets of sizes from $1$ to $\min(km,n)$. 
\end{itemize}

In one individual stage of $U_{n,k,m}$ we have $m!$ orderings of a single allocation of $m$-subset of one step of $U_{n,k,\left\{m\right\}}$. This is to be taken into account comparing the schemes $U_{n,k,\left\{m\right\}}$ and $U_{n,k,m}$. This difference can also be seen comparing the one stage outcomes of $U_{n,k,\left\{m\right\}}$ and $U_{n,k,m}$. Represent $C_{n}^{m}$ of model $U_{n,k,\left\{m\right\}}$ as 
\begin{align}
\frac{n!}{m!(n-m)!}=\frac{n(n-1) \ldots (n-m+1)}{m!}.
\end{align}
Numerator of the last ratio is the counterpart of $n^{m}$ of model $U_{n,k,\left\{m\right\}}$, and $m!$ is the coefficient we mentioned about. Comparing $U_{n,k,\left\{m\right\}}$ and $U_{n,k,m}$, first we note that outcomes of $U_{n,k,\left\{m\right\}}$ are part of outcomes of $U_{n,k,m}$ and hence they may have higher probabilities.
\begin{figure}[tb]
\begin{center}
\begin{minipage}[h]{\linewidth}  
\includegraphics[width=.9\textwidth]{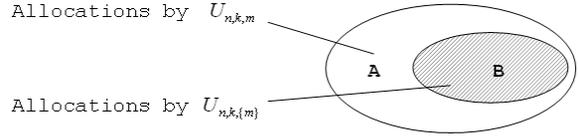}
\end{minipage}
\end{center}
\caption{Allocations by $U_{n,k,\left\{m\right\}}$ and $U_{n,k,m}$}
\label{fig:allocations}
\end{figure}
Consider the probability $p_{j}$ of an event, that in stage $j$ of $U_{n,k,m}$, all the allocated $m$ elements are different. Then $P=p_{1} \cdot p_{2} \cdot \ldots \cdot p_{k}$ is the probability that in all $k$ stages allocated $m$ elements are different. In different stages allocations of course may intersect. Outcomes of $U_{n,k,\left\{m\right\}}$ multiplied with this probabilities are equal to probabilities of $U_{n,k,m}$, part $B$ of intersection of outcomes (Fig. \ref{fig:allocations}). $p_{j}$ Was estimated in previous point as a value tending to $1$ asymptotically. We may extend this proposition to the entire value $P$.
Formally we use the property that probability of union of events is less or equal the sum of event probabilities:

\begin{align}
Pr & \left\{(v_{n,m} \geq \epsilon | q=1) \vee \ldots \vee (v_{n,m} \geq \epsilon | q=k) \right\} \leq \nonumber \\
&\leq k \cdot Pr\left\{v_{n,m} \geq \epsilon \right\} \leq \frac{k \cdot M(v_{n,m})}{\epsilon}.
\end{align}

Then the final condition (upper estimate) sufficient for repetition probability tending to zero is $kC_{m}^{2}/n\rightarrow 0$ with $n,m,k\rightarrow \infty$. The sufficient condition for allocation of all $m$ agents in all $k$ consecutive stages to different nodes $km^{2}/n\rightarrow 0$ is naturally acceptable in WSN which have a very large nodes set as a rule.
Final picture is: part $B$ allocations (Fig. \ref{fig:allocations}) appear in $U_{n,k,m}$ with probability $P$ tending to $1$; relative probability distribution among the elements of $B$ is identical in $U_{n,k,\left\{m\right\}}$ and $U_{n,k,m}$; event probability in model $U_{n,k,\left\{m\right\}}$ is not less than in $U_{n,k,m}$ multiplied by $P$; probabilities of  $t$-subset allocations under the model $U_{n,k,m}$ have formulas similar to the ones for model $U_{n,k,\left\{m\right\}}$ considered above.

If $R(k,m,t)$ denotes the number of $t$-node allocations in model $U_{n,k,m}$ then the formal representation of $R(k,m,t)$ similar to the formula for $Q(k,m,t)$. Considered above can be achieved by the same inclusion exclusion method:
\begin{align}
R&(k,m,t)=t^{mk} - C_{t}^{1} \cdot (t-1)^{mk} + C_{t}^{2} \cdot (t-2)^{mk} - \ldots \nonumber \\
&\ldots + (-1)^{t-1}C_{t}^{t-1} \cdot (t-(t-1))^{mk} = \nonumber \\ &=\sum_{i=0}^{t-1}(-1)^{i}C_{t}^{i} \cdot (t-i)^{mk}. \label{rkmtformula}
\end{align}
On this basis we formulate

\begin{theorem}
If $kC_{m}^{2}/n\rightarrow 0$ with $n,m,k\rightarrow \infty$, then comparison of $U_{n,k,\left\{m\right\}}$ and $U_{n,k,m}$ model probabilities of  $t$-node allocations are by relation
\begin{align}
	\frac{C_{n}^{t}Q(k,m,t)}{(C_{n}^{m})^k} \cdot P \leq \frac{C_{n}^{t}R(k,m,t)}{n^{km}} \text{, with }  P \rightarrow 1.
\end{align}
\end{theorem}
Finally, we note that $R(k,m,t)$ has equivalent presentation in terms of second kind Stirling numbers (\cite{chelluri}) 
\begin{align}
S(N,K) = \frac{1}{K!} \sum_{j=0}^{K}(-1)^{j}C_{K}^{j} (K-j)^N.
\end{align}
Here we used the fact that allocation of $k$ consecutive and independent stages of $m$ elements over the WSN area of $n$ nodes is equivalent to allocation of $km$ elements over that area. Note a difference between the formulas for $Q(k,m,t)$ and $R(k,m,t)$ - that is summation limits. In case of $R(k,m,t)$ formally we may add the zero term for $i=t$, and then we receive
\begin{align}
	R(k,m,t)=t!S(mk,t)
\end{align}
which is the final postulation of this paper.

\section{Conclusion}
WSN and software agent systems are important application technique for many areas. Being hard algorithmically and complex in model level these systems require special economy regimes and this is concerned in knowing the minimal requirements and maximum effect when resource is limited. In randomly roving agents model, which is considered above, it is shown that appearing probabilities are equivalently presented in terms of combinatorial Stirling numbers and due to known asymptotic formulas for these numbers (\cite{chelluri}), this allows to adopt the monitoring regime in an optimal way.

\bibliographystyle{plain}
\bibliography{biblo}
\end{document}